\documentclass{article}

\usepackage{microtype}
\usepackage{graphicx}
\usepackage{subcaption}
\usepackage{booktabs} 

\usepackage{hyperref}

\usepackage[preprint]{icml2026}

\usepackage{amsmath}
\usepackage{amssymb}
\usepackage{mathtools}
\usepackage{amsthm}
\usepackage{enumitem}

\usepackage[capitalize,noabbrev]{cleveref}
\usepackage{amsmath, amssymb}
\usepackage{booktabs}
\usepackage{array}

\newcommand{\R}{\mathbb{R}}
\newcommand{\study}{\mathcal{S}}
\newcommand{\viewset}{\mathcal{V}}

\theoremstyle{plain}

\theoremstyle{definition}

\theoremstyle{remark}

\usepackage[disable,textsize=tiny]{todonotes}

\icmltitlerunning{EchoJEPA: A Latent Predictive Foundation Model for Echocardiography}

\begin{document}

\twocolumn[
  \icmltitle{EchoJEPA: A Latent Predictive Foundation Model for Echocardiography}




  \icmlsetsymbol{equal}{*}
  \icmlsetsymbol{senior}{\textdagger}

  \begin{icmlauthorlist}
    \icmlauthor{Alif Munim}{equal,uhn,cohere}
    \icmlauthor{Adibvafa Fallahpour}{equal,uhn,vector,uoft}
    \icmlauthor{Teodora Szasz}{equal,uofc,philips}
    \icmlauthor{Ahmadreza Attarpour}{equal,uhn}
    \icmlauthor{River Jiang}{ucsf}
    \icmlauthor{Brana Sooriyakanthan}{uhn}
    \icmlauthor{Maala Sooriyakanthan}{uhn}
    \icmlauthor{Heather Whitney}{uofc}
    \icmlauthor{Jeremy Slivnick}{uofc}
    \icmlauthor{Barry Rubin}{uhn,uoft}
    \icmlauthor{Wendy Tsang}{senior,uhn,uoft}
    \icmlauthor{Bo Wang}{senior,uhn,vector,uoft}
  \end{icmlauthorlist}

  \icmlaffiliation{uhn}{University Health Network}
  \icmlaffiliation{uoft}{University of Toronto}
  \icmlaffiliation{uofc}{University of Chicago}
  \icmlaffiliation{ucsf}{University of California, San Francisco}
  \icmlaffiliation{vector}{Vector Institute}
  \icmlaffiliation{cohere}{Cohere Labs}
  \icmlaffiliation{philips}{Philips Health}

    \icmlcorrespondingauthor{Alif Munim}{alif.munim@uhn.ca}
    \icmlcorrespondingauthor{Wendy Tsang}{wendy.tsang@uhn.ca}
    \icmlcorrespondingauthor{Bo Wang}{Bo.Wang@uhn.ca}

  \icmlkeywords{Machine Learning, ICML}

  \vskip 0.3in
]



\printAffiliationsAndNotice{\icmlEqualContribution \textsuperscript{\textdagger}Co-senior authors.}

\begin{abstract}
Foundation models for echocardiography often struggle to disentangle anatomical signal from the stochastic speckle and acquisition artifacts inherent to ultrasound. We present EchoJEPA, a foundation model trained on 18 million echocardiograms across 300K patients, representing the largest pretraining corpus for this modality to date. By leveraging a latent predictive objective, EchoJEPA learns robust anatomical representations that ignore speckle noise. We validate this using a novel multi-view probing framework with frozen backbones, where EchoJEPA outperforms leading baselines by approximately 20\% in left ventricular ejection fraction (LVEF) estimation and 17\% in right ventricular systolic pressure (RVSP) estimation. The model also exhibits remarkable sample efficiency, reaching 79\% view classification accuracy with only 1\% of labeled data versus 42\% for the best baseline trained on 100\%. Crucially, EchoJEPA demonstrates superior generalization, degrading by only 2\% under physics-informed acoustic perturbations compared to 17\% for competitors. Most remarkably, its zero-shot performance on pediatric patients surpasses fully fine-tuned baselines, establishing latent prediction as a superior paradigm for robust, generalizable medical AI.

\end{abstract}

\section{Introduction}
\label{sec:intro}

Echocardiography is the most widely used cardiac imaging modality, with approximately 30 million studies performed annually in the United States alone \citep{Virnig_Shippee_O’Donnell_Zeglin_Parashuram_2011}. Its accessibility and lack of ionizing radiation make it the first-line tool for evaluating cardiac structure and function \citep{Lang2015}. Recent efforts have sought to develop foundation models for echocardiography that learn generalizable representations from large unlabeled video archives, promising to reduce annotation burden and improve diagnostic consistency \citep{panecho, echoprime, adibi2025recentadvancesapplicationsopen}. However, these models face challenges arising from the unique signal properties of ultrasound that distinguish it from natural video domains.

Ultrasound video is dominated by stochastic speckle patterns, depth-dependent intensity attenuation, and acoustic shadows, artifacts that vary across acquisitions and bear no relationship to cardiac anatomy \citep{1539054}. Existing foundation models use supervised multitask learning \citep{panecho}, contrastive vision-language alignment \citep{echoprime}, or masked autoencoding \citep{kim2025echofmfoundationmodelgeneralizable, mae, videomae}, yet none explicitly targets noise-invariant representations: supervised models inherit annotation noise, contrastive models align to report language rather than anatomy, and reconstruction models must faithfully reproduce speckle to minimize their loss.

We demonstrate that latent prediction provides a superior objective. Joint-embedding predictive architectures (JEPA) train a predictor to infer embeddings of masked regions from visible context, targeting an exponential moving average teacher rather than raw pixels \citep{vjepa, vjepa2, chen2025vljepajointembeddingpredictive}. This downweights unpredictable artifacts like speckle while reinforcing temporally coherent structures like chamber geometry and wall motion.

We introduce \textbf{EchoJEPA}, the first foundation-scale application of joint-embedding predictive architectures to echocardiography, pretrained on 18 million videos across 300K patients. EchoJEPA adapts V-JEPA2 with domain-appropriate temporal resolution and augmentation \citep{vjepa2}. To handle variable study composition, we develop a multi-view probing framework with factorized video stream embeddings that integrates information across views without view-specific components, offering a superior alternative to prior multi-view embedding scheme \citep{Tohyama2025.08.15.25333725}. 

\paragraph{Contributions.} Our key contributions are as follows:

\begin{itemize}[leftmargin=10pt, itemsep=2pt, topsep=0pt]
\item \textbf{EchoJEPA.} A foundation model using latent prediction pretrained on 18 million videos across 300K patients, the largest echocardiography corpus to date, achieving state-of-the-art performance on left ventricular ejection fraction (LVEF) estimation and right ventricular systolic pressure (RVSP) prediction, demonstrating that latent prediction outperforms pixel reconstruction for ultrasound.
\item \textbf{Multi-view probing framework.} A method using factorized video stream embeddings and attention masking to integrate information across echocardiographic views without view-specific components.
\item \textbf{Unified evaluation protocol.} A standardized benchmark with frozen backbones, identical probes, and consistent hyperparameter search across all baseline models, enabling fair comparison of representation quality.
\item \textbf{Robustness benchmarks.} Physics-informed perturbations using depth attenuation and acoustic shadow \citep{pmlr-v172-singla22a}, revealing that EchoJEPA degrades 86\% less than the next-best baseline under acoustic perturbations.
\item \textbf{Public release.} We open-source EchoJEPA-L, a state-of-the-art echocardiography foundation model trained on MIMIC-IV-Echo \citep{PhysioNet-mimic-iv-echo-0.1}, alongside our evaluation framework at \url{https://github.com/bowang-lab/EchoJEPA}.
\end{itemize}

    
    
    
    



\section{Related Work}
\label{sec:related}

\subsection{Self-Supervised Video Learning}
\label{sec:related:ssl}
Self-supervised video representation learning follows two dominant paradigms \citep{han2021selfsupervisedcotrainingvideorepresentation}. Reconstruction-based methods learn by imputing masked spatiotemporal content in pixel space. Masked Autoencoders \cite{mae} demonstrated that high masking ratios with lightweight decoders enable scalable pretraining, and VideoMAE \cite{videomae} and ST-MAE \cite{stmae} extended this to video by masking space-time tubelets. Prediction-based methods instead target learned representations rather than raw pixels. I-JEPA \cite{ijepa} predicts embeddings of masked image regions from visible context, while V-JEPA \cite{vjepa} and V-JEPA2 \cite{vjepa2} extend this to video using an EMA target encoder for stable prediction targets. Notably, pixels reward fidelity to surface statistics while embeddings reward capture of semantically stable structure \citep{mishra2026selfsupervisedlearningechocardiographicvideo}. We test which paradigm suits domains where pixel-level fidelity is dominated by stochastic nuisance factors, a regime unexplored by prior work on natural video.

\subsection{Foundation Models for Echocardiography}
\label{sec:related:echo}
Building on foundation model success in healthcare, echocardiography has followed suit \citep{medsam,medsam2,fallahpour2024ehrmambageneralizablescalablefoundation}. EchoNet-Dynamic \cite{echonet} released 10,030 annotated apical-4-chamber videos and established supervised baselines for ejection fraction estimation, and EchoNet-Pediatric \cite{echonet_pediatric} extended this to pediatric populations. Recent foundation models pursue broader task coverage through self-supervised pretraining. PanEcho \cite{panecho} trains a ConvNeXt encoder with temporal transformer on over one million videos, achieving view-agnostic prediction through post-hoc averaging of clip-level outputs. EchoPrime \cite{echoprime} scales to 12 million video-report pairs with contrastive vision-language learning, using view classification and attention-based pooling for study-level predictions. Similarly, EchoCLIP \citep{echoclip} aligns visual representations with text reports via contrastive learning on 1 million video-text pairs. EchoFM applies masked autoencoding to 290,000 videos to learn spatiotemporal representations \citep{kim2025echofmfoundationmodelgeneralizable}.

Distinct from diagnostic foundation models, EchoWorld \cite{echoworld} applies world modeling to robotic probe guidance, integrating motion-aware attention to predict visual dynamics conditional on 6-DOF probe pose data. In the low-data regime, Ellis et al. \cite{ellis2025} utilized V-JEPA for segmentation on the small CAMUS dataset \citep{Leclerc_2019}, introducing auxiliary localization tasks to compensate for the lack of inductive bias in Vision Transformers \citep{dosovitskiy2021imageworth16x16words}. Unlike these approaches, we scale latent prediction to 18 million videos to build a general-purpose diagnostic encoder without reliance on hardware pose data or auxiliary losses.

\subsection{Robustness Evaluation in Medical Imaging}
\label{sec:related:robust}
Standard evaluation protocols emphasize i.i.d. test performance, underestimating failure modes in clinical deployment where distribution shift is ubiquitous \cite{medical_domain_shift}. Echocardiography exhibits acquisition variability since image quality depends on patient body habitus, probe positioning, operator expertise, and equipment vendor \cite{echo_variability}. Patients most likely to benefit from automated analysis, such as those with obesity or limited acoustic windows, are those whose images deviate most from training distributions, yet prior robustness evaluations focus on ImageNet-C corruptions \cite{imagenetc} or adversarial perturbations \cite{medical_adversarial}, neither capturing ultrasound-specific degradation.



\begin{figure*}[t!]
    \centering
    \includegraphics[width=\linewidth]{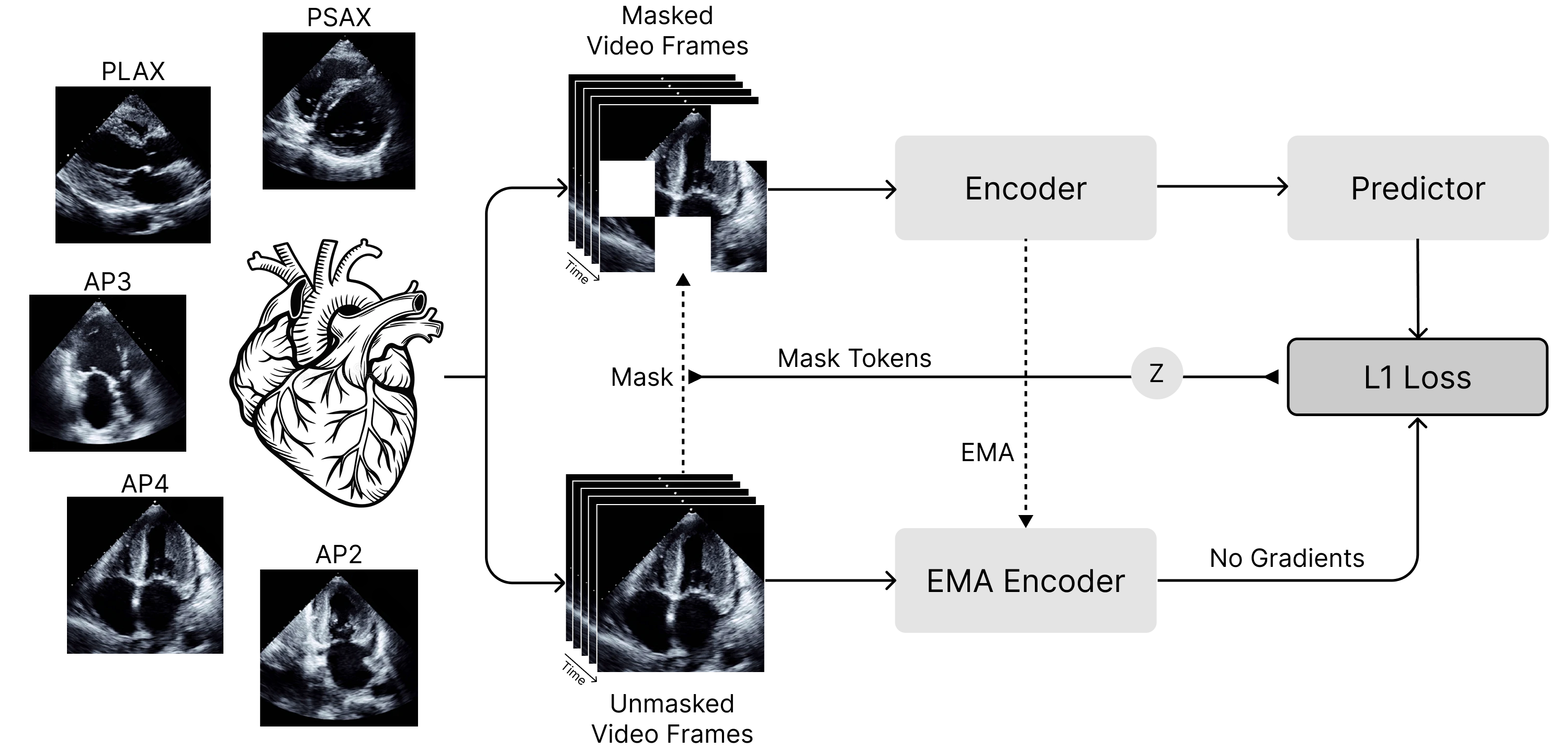}
    \caption{
        \textbf{EchoJEPA architecture.}
        Echocardiograms from multiple views are partitioned into spatio-temporal tubelets and split into masked and unmasked sets.
        The encoder $E_\theta$ processes visible (unmasked) video frames, and the predictor $P_\phi$ infers embeddings for masked regions conditioned on learnable mask tokens.
        The EMA encoder $E_{\bar{\theta}}$ processes unmasked frames to provide prediction targets.
        The $L_1$ loss is computed between predicted and target embeddings, with no gradients flowing into the EMA encoder.
    }
    \label{fig:method1a}
\end{figure*}

\section{EchoJEPA}
\label{sec:method}

We introduce EchoJEPA, the first foundation-scale application of joint-embedding predictive architectures to echocardiography. Trained on 18 million videos, EchoJEPA builds on V-JEPA2 \citep{vjepa2}, which learns video representations by predicting masked spatio-temporal content in embedding space rather than pixel space (\cref{fig:method1a}).


\subsection{Latent Predictive Pretraining}
\label{sec:method:jepa}

Given an input video $V = \{t_1, t_2, \ldots, t_N\}$ partitioned into $N$ spatio-temporal tubelets, which are 3D patches spanning multiple frames and spatial regions, we divide these into two disjoint sets: context tubelets $x$ that remain visible and target tubelets $y$ that are masked.

The architecture comprises three components. The context encoder $E_\theta$ extracts representations from visible tubelets. The predictor $P_\phi$ takes these representations along with a learnable mask token $\Delta_y$ indicating the spatio-temporal positions of masked tubelets and infers embeddings for the masked regions. The target encoder $E_{\bar{\theta}}$ produces the prediction targets. The model minimizes the $L_1$ distance between predicted and target embeddings:

\begin{equation}
    \mathcal{L} = \left\| P_\phi(\Delta_y, E_\theta(x)) - sg(E_{\bar{\theta}}(y)) \right\|_1
    \label{eq:jepa_loss}
\end{equation}

The stop-gradient operator $sg(\cdot)$ prevents gradients from flowing into the target encoder, which would otherwise cause representational collapse. Instead, the target encoder weights are updated as an exponential moving average of the context encoder after each training step, ensuring prediction targets remain stable yet adapt gradually during training.

This objective naturally suits echocardiography because the slowly evolving target encoder suppresses stochastic speckle while reinforcing spatio-temporally coherent structures such as chamber geometry and wall motion.




\subsection{Domain Adaptations for Echocardiography}
\label{sec:method:domain}





\begin{figure*}[t!]
\centering
\includegraphics[width=\linewidth]{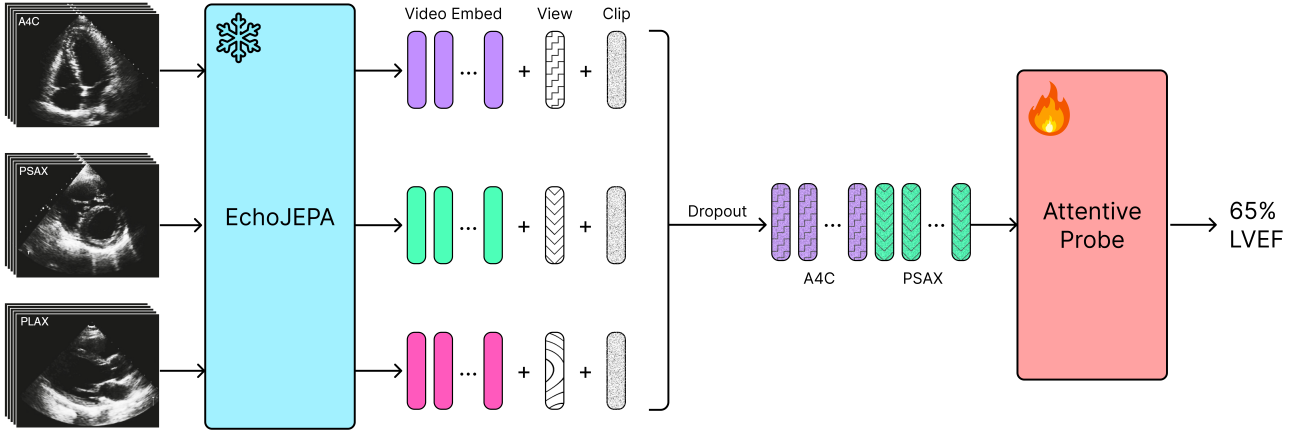}
\caption{
\textbf{Multi-view probing framework.}
The frozen EchoJEPA encoder extracts video embeddings from multiple echocardiographic views. Each embedding is augmented with learnable view and clip stream embeddings encoding position in the study. During training, view dropout randomly masks views to improve robustness to variable study composition. The concatenated tokens are passed to a lightweight attentive probe that outputs study-level predictions.
}
\label{fig:multiview}
\end{figure*}

We adapt V-JEPA2 to echocardiography with three modifications that respect the signal properties of ultrasound.

\paragraph{Temporal resolution.}
We increase the sampling rate from 4 fps to 24 fps. Cardiac dynamics unfold rapidly, some within 50–100 ms, requiring higher temporal resolution to capture sufficient frames.

\paragraph{Aspect ratio augmentation.}
We narrow the random aspect ratio range from $(0.75, 1.35)$ to $(0.9, 1.1)$. Echocardiographic views follow standardized acquisition protocols with consistent geometry, and aggressive augmentation distorts clinically meaningful chamber proportions.

\paragraph{Crop scale augmentation.}
We adjust the random crop scale range from $(0.3, 1.0)$ to $(0.5, 1.0)$. The ultrasound sector has a fan-shaped geometry, and crops below 50\% risk excluding cardiac structures entirely.

\subsection{Model Variants and Training}
\label{sec:method:models}

We instantiate EchoJEPA at two scales to study the interplay between model capacity, training data, and representation quality (Table~\ref{tab:models}).

Our flagship model uses ViT-Giant (1.1B parameters) pretrained on 18.1M proprietary echocardiogram videos spanning diverse populations and scanner manufacturers \citep{dosovitskiy2021imageworth16x16words}. Training proceeds in two phases: pretraining at $224^2$ resolution for 280 epochs, followed by annealing at $336^2$ for 80 epochs with reduced learning rate. For reproducibility, we provide EchoJEPA-L using ViT-Large (300M parameters) trained on MIMIC-IV-Echo \citep{PhysioNet-mimic-iv-echo-0.1} (525K public videos), enabling external validation of our methodology.

To isolate pretraining objective from confounding factors, we train VideoMAE \citep{videomae} with identical architecture, data, augmentations, and compute to test whether latent prediction confers advantages for ultrasound independent of other factors.

\begin{table}[b]
\centering
\caption{Model configurations. $^\dagger$Compute-matched baseline for objective comparison.}
\label{tab:models}
\vspace{0.5em}
\small
\begin{tabular}{@{}lccccc@{}}
\toprule
\textbf{Model} & \textbf{Backbone} & \textbf{Training Data} & \textbf{Videos} \\
\midrule
EchoJEPA-G & ViT-G & Proprietary &  18.1M \\
EchoJEPA-L & ViT-L & MIMIC-IV-Echo & 525K \\
VideoMAE-L$^\dagger$ & ViT-L & MIMIC-IV-Echo & 525K  \\
\bottomrule
\end{tabular}
\end{table}

\subsection{Multi-View Attentive Probing}
\label{sec:method:probing}

Existing echocardiography foundation models employ heterogeneous evaluation strategies that confound representation quality with architectural choices \citep{panecho, echoprime}. We introduce a standardized probing framework that fixes the probe architecture, hyperparameter search, and multi-view fusion strategy across all models, isolating representation quality as the sole variable. The overall pipeline is depicted in Figure~\ref{fig:multiview}.

\paragraph{Study structure.}
An echocardiography study $\study = \{v_1, \ldots, v_N\}$ contains $N$ video clips with associated view labels. For a clinical task with relevant view subset $\viewset_{\text{task}}$ containing $V$ views, we select one video per view. For each video, we sample $C$ temporal clips, yielding $L = V \times C$ streams. A binary mask $\mathbf{m} \in \{0,1\}^V$ indicates view availability. During training, we dropout views randomly with probability $p_{\text{miss}} = 0.1$ for robustness to incomplete studies.

\paragraph{Architecture.}
Each frozen encoder $E$ maps a clip to $N_E$ tokens of dimension $D$, producing $\{\mathbf{e}^{(\ell)} \in \R^{N_E \times D}\}_{\ell=1}^L$ across all streams. We concatenate these into $\mathbf{X} = [\mathbf{e}^{(1)}; \ldots; \mathbf{e}^{(L)}] \in \R^{(L \cdot N_E) \times D}$. To encode stream identity, we introduce factorized learnable embeddings $\mathbf{E}^{\text{view}} \in \R^{V \times D}$ and $\mathbf{E}^{\text{clip}} \in \R^{C \times D}$, requiring only $(V + C) \times D$ parameters versus $L \times D$ for full stream embeddings. For each token $i$ belonging to stream $\ell$ with view index $v_\ell$ and clip index $c_\ell$:
\begin{equation}
    \tilde{\mathbf{X}}_i = \mathbf{X}_i + \mathbf{E}^{\text{view}}_{v_\ell} + \mathbf{E}^{\text{clip}}_{c_\ell}
\end{equation}
Tokens pass through $R-1$ self-attention blocks with a key padding mask $\mathbf{K}$ derived from $\mathbf{m}$ that ignores tokens from missing views. A learnable query $\mathbf{q} \in \R^{1 \times D}$ then cross-attends to the output tokens $\mathbf{X}' \in \R^{(L \cdot N_E) \times D}$ to produce the final representation \citep{vaswani2023attentionneed}:
\begin{equation}
    \mathbf{h} = \text{CrossAttn}(\mathbf{q}, \mathbf{X}', \mathbf{K}) \in \R^{D}
\end{equation}
A linear head maps $\mathbf{h}$ to the task prediction.

\paragraph{Protocol.}
All models use identical probes with depth $R = 4$, 16 attention heads, and MLP ratio 4. Following V-JEPA 2 \citep{vjepa2}, we train across a hyperparameter grid over learning rates $\eta \in \{10^{-4}, 5 \times 10^{-5}\}$ and weight decay $\lambda \in \{0.01, 0.1, 0.4\}$, reporting best performance for each model. This standardization ensures differences reflect representation quality rather than probe design.

\subsection{Robustness Evaluation Protocol}
\label{sec:method:robustness}

Standard i.i.d.\ evaluation underestimates failure modes in clinical deployment where distribution shift is common \citep{oakdenrayner2020hidden}. We introduce physics-informed perturbations simulating the dominant degradation modes in echocardiography \citep{tupper2025}.

\textbf{Depth attenuation.} Ultrasound signal intensity decreases with 
tissue depth due to absorption and scattering, particularly in patients 
with obesity or poor acoustic windows. We simulate this by applying a 
linear intensity ramp:
\begin{equation}
    I'(x, y) = I(x, y) \cdot \max\left(0, 1 - \alpha \cdot \frac{y}{H}\right)
\end{equation}
where $I(x,y)$ is the original pixel intensity, $I'(x,y)$ is the 
perturbed intensity, $x$ and $y$ are horizontal and vertical pixel 
coordinates, $H$ is image height, and $\alpha \in \{0.3, 0.5, 0.7\}$ 
controls attenuation severity.

\textbf{Acoustic shadow.} Shadows occur when ultrasound is blocked by 
highly reflective structures such as ribs or calcifications. We simulate 
this with a Gaussian-weighted intensity reduction along the horizontal axis:
\begin{equation}
    I'(x, y) = I(x, y) \cdot \left(1 - \exp\left(-\frac{(x - x_0)^2}{2\sigma^2}\right)\right)
\end{equation}
where $x_0$ is the shadow center sampled uniformly across the image width, 
and $\sigma \in \{0.1W, 0.2W, 0.3W\}$ controls shadow width relative to 
image width $W$. Both perturbations are applied consistently across all 
frames to simulate realistic acquisition conditions.

\begin{figure}[t!]
    \centering
    \includegraphics[width=\linewidth]{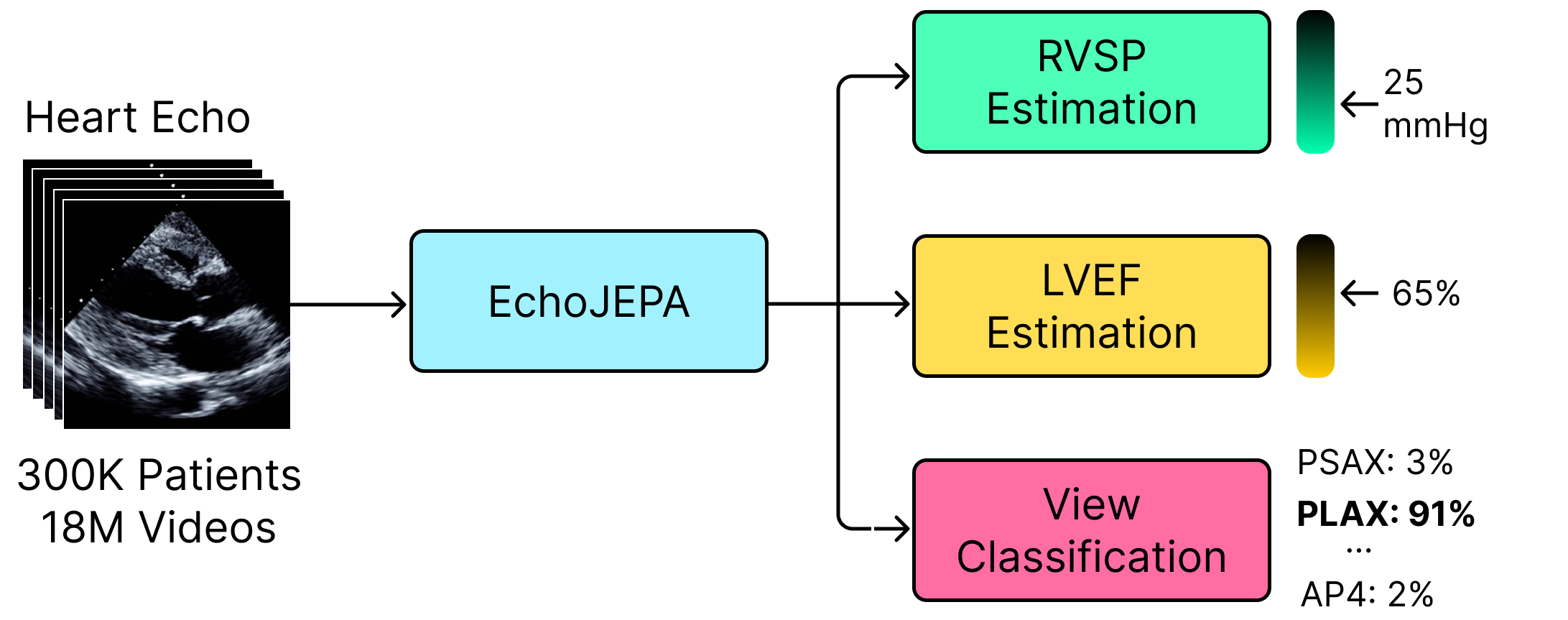}
    \caption{
        \textbf{Downstream evaluation.} EchoJEPA pretrained on 300K patients and 18M videos is evaluated on three clinical tasks with frozen backbones and lightweight probes. RVSP estimation, LVEF regression, and view classification.
    }
    \label{fig:method1b}
\end{figure}

\section{Experiments}
\label{sec:experiments}

We evaluate EchoJEPA on three axes that determine clinical utility: (1) whether latent prediction outperforms reconstruction under controlled conditions, (2) sample efficiency and robustness under distribution shift, and (3) generalization across patient populations and multi-view reasoning tasks.

\subsection{Experimental Setup}
\label{sec:exp:setup}

\paragraph{Datasets.}
We utilize two large-scale internal health networks and two public benchmarks to assess generalization:
\begin{itemize}[leftmargin=*, nosep, topsep=0pt]

\item \textbf{Toronto (Internal)}: $N{=}150,000$ studies used for probe training and internal validation.

\item \textbf{Chicago (Internal)}: $N{=}60,000$ studies used as an external holdout site.

\item \textbf{EchoNet-Dynamic}~\citep{echonet}: 10,030 videos (Stanford) used for external zero-shot evaluation of LVEF.

\item \textbf{EchoNet-Pediatric}~\citep{echonet_pediatric}: 3,316 videos used to test generalization across patient populations.

\end{itemize}

\paragraph{Tasks.}
We evaluate on three tasks representing the clinical pipeline from triage to diagnosis (\cref{fig:method1b}):
\begin{itemize}[leftmargin=*, nosep, topsep=0pt]

\item \textbf{View Classification}: 12-class identification of standard views (Accuracy \%). Serves as a triage primitive for automated pipelines.

\item \textbf{LVEF Regression}: Left ventricular ejection fraction from apical views (MAE \%, lower is better). The standard metric for cardiac function.

\item \textbf{RVSP Regression}: Right ventricular systolic pressure from multi-view integration (MAE mmHg, lower is better). Requires reasoning across apical (TR velocity) and subcostal (IVC diameter) views.

\end{itemize}

\paragraph{Models.}
We compare five foundation models with different objectives and scales:
\begin{itemize}[leftmargin=*, nosep, topsep=0pt]

\item \textbf{EchoJEPA-G}: ViT-Giant (1.1B params) trained with latent prediction on 18M proprietary echocardiograms.

\item \textbf{EchoJEPA-L}: ViT-Large (300M params) trained with latent prediction on 525K videos from MIMIC-IV-Echo~\citep{PhysioNet-mimic-iv-echo-0.1} (public).

\item \textbf{EchoMAE-L}: ViT-Large trained with pixel reconstruction (VideoMAE objective) on MIMIC-IV-Echo~\citep{PhysioNet-mimic-iv-echo-0.1}. This model is compute-matched to EchoJEPA-L, differing only in objective.

\item \textbf{PanEcho}~\citep{panecho}: ConvNeXt-Tiny (28M params) trained end-to-end on 1M+ videos with supervised multitask learning across 39 clinical outputs.

\item \textbf{EchoPrime}~\citep{echoprime}: mViT-v2 (35M params) trained on 12M video-report pairs with contrastive learning.

\end{itemize}

\paragraph{Unified Evaluation Protocol.}
Prior works evaluate with heterogeneous protocols such as fine-tuning versus linear probing or averaging versus retrieval, confounding representation quality with adaptation strategy.
To ensure rigorous comparability, we enforce a strictly unified protocol:
\begin{itemize}[leftmargin=*, nosep, topsep=0pt]

\item \textbf{Standardized Encoder Wrappers}: We implement wrappers for PanEcho~\citep{panecho} and EchoPrime~\citep{echoprime} that align their outputs with the attentive probe interface.

\item \textbf{Frozen Backbones}: All encoder weights are frozen; only probe parameters are trained.

\item \textbf{Identical Probes}: All models use the same attentive probe~\citep{vjepa2} (depth=4, 16 heads).

\item \textbf{Identical Hyperparameters}: All models undergo identical learning rate and weight decay sweeps.

\item \textbf{Same Multi-View Fusion}: We use our early fusion framework with stream embeddings for all models on multi-view tasks.

\end{itemize}

\subsection{Controlled Comparison: Latent vs. Pixel Prediction}
\label{sec:exp:controlled}

We isolate the effect of pretraining objective by comparing EchoJEPA-L and EchoMAE-L, which share identical architecture (ViT-L), training data (MIMIC-IV-Echo, 525K videos), augmentations, and compute budget. The sole difference is the objective: latent prediction (L1 loss on EMA encoder targets) versus pixel reconstruction (MSE loss on masked patches). 

We also compare EchoJEPA to EchoPrime and PanEcho, which use different objectives (contrastive VLM and supervised multitask, respectively) and substantially more training data (12M and 1M+ videos vs. 525K for EchoJEPA-L). We include these comparisons alongside the compute-matched pretraining comparison in Table~\ref{tab:controlled}.

\begin{table}[t]
    \centering
    \caption{
        \textbf{Controlled comparison of pretraining objectives.}
        EchoJEPA-L and EchoMAE-L use identical architecture, data, and compute.
        Latent prediction consistently outperforms pixel reconstruction.
    }
    \label{tab:controlled}
    \vspace{0.5em}
    \small
    \begin{tabular}{@{}llcc@{}}
        \toprule
        \textbf{Model} & \textbf{Objective} & \textbf{LVEF MAE $\downarrow$} & \textbf{View Acc $\uparrow$} \\
        \midrule
        EchoMAE-L & Reconstruction & 8.15 & 40.4 \\
        EchoJEPA-L & Latent Prediction & \textbf{5.97} & \textbf{85.5} \\
        \midrule
        \multicolumn{2}{@{}l}{\textit{Relative improvement}} & \textit{-\textbf{26.7}\%} & \textit{+\textbf{45.1}\%} \\
        \bottomrule
    \end{tabular}
\end{table}

Under identical conditions, EchoJEPA-L outperforms EchoMAE-L by 26.7\% on LVEF estimation and 45.1\% on view classification. This result confirms that latent prediction offers superior performance over pixel reconstruction. Table~\ref{tab:lvef} compares all models on LVEF estimation. EchoJEPA-G achieves 4.26 MAE on Toronto (vs.\ 5.33 for EchoPrime) and 3.97 MAE on Stanford (vs.\ 4.87 for EchoPrime).

\begin{table}[t]
    \centering
    \caption{\textbf{LVEF estimation} (MAE, lower is better). All models 
    evaluated with frozen backbones and identical attentive probes. Probes 
    trained on Toronto; Chicago results demonstrate cross-site generalization.}
    \label{tab:lvef}
    \vspace{0.5em}
    \small
    \begin{tabular}{@{}lccc@{}}
        \toprule
        \textbf{Model} & \textbf{Toronto} & \textbf{Chicago} & \textbf{Stanford}$^{\dagger}$ \\
        \midrule
        EchoPrime & 5.33 & 6.71 & 4.87 \\
        PanEcho & 5.43 & 6.52 & 5.10 \\
        \midrule
        EchoMAE-L & 8.15 & 9.40 & 8.52 \\
        EchoJEPA-L & 5.97 & 7.39 & 5.76 \\
        EchoJEPA-G & \textbf{4.26} & \textbf{5.44} & \textbf{3.97} \\
        \bottomrule
    \end{tabular}
    \begin{minipage}{\linewidth}
        \centering
        \scriptsize $^\dagger$ Probes trained and evaluated on the public EchoNet-Dynamic dataset splits.
    \end{minipage}
\end{table}

\subsection{Sample Efficiency}
\label{sec:exp:efficiency}

Table~\ref{tab:sample_efficiency} demonstrates that EchoJEPA models trained on just 1\% of labeled data outperform baselines trained on 100\%. Even the publicly trained EchoJEPA-L achieves 57.6\% accuracy with 1\% labels (vs.\ 42.1\% for EchoPrime), while EchoJEPA-G reaches 78.6\%, nearly double the fully-supervised baseline. This efficiency implies that latent prediction yields dense representations capable of defining the view manifold with minimal supervision, as evidenced by the distinct anatomical clustering in Figure~\ref{fig:umap_views}.

\begin{table}[t]
    \centering
\caption{\textbf{Sample efficiency on view classification} (Accuracy \%). 
EchoJEPA at 1\% labels outperforms all baselines trained on 100\%. Values represent mean $\pm$ standard deviation over 3 runs.}
    \label{tab:sample_efficiency}
    \vspace{0.5em}
    \small
    \begin{tabular}{@{}lccc@{}}
        \toprule
        \textbf{Model} & \textbf{1\%} & \textbf{10\%} & \textbf{100\%} \\
        \midrule
        EchoPrime & 21.63$_{\pm 0.55}$ & 32.06$_{\pm 0.81}$ & 42.1 \\
        PanEcho & 21.48$_{\pm 0.60}$ & 30.62$_{\pm 0.15}$ & 41.9 \\
        \midrule
        EchoMAE-L & 21.86$_{\pm 1.26}$ & 34.47$_{\pm 1.22}$ & 40.4 \\
        EchoJEPA-L & 57.55$_{\pm 0.72}$ & 80.06$_{\pm 0.87}$ & 85.5 \\
        EchoJEPA-G & \textbf{78.63}$_{\pm 1.21}$ & \textbf{84.42}$_{\pm 0.14}$ & \textbf{87.4} \\
        \bottomrule
    \end{tabular}
\end{table}

This striking sample efficiency suggests that latent prediction yields semantically dense representations where a tiny fraction of labels suffices to define the view manifold. In contrast, baseline models struggle to separate acquisition noise from semantic signal even with two orders of magnitude more labeled data.

\subsection{Robustness to Acoustic Degradation}
\label{sec:exp:robustness}

We evaluate performance under physics-informed perturbations: depth attenuation and Gaussian shadow (Table~\ref{tab:robustness_full}).

\begin{table*}[t]
    \centering
    \caption{
        \textbf{Robustness to acoustic degradation} (LVEF MAE on Stanford, lower is better).
        EchoJEPA degrades more gracefully than alternative foundation models under depth attenuation and Gaussian shadowing.
        Avg.\ Deg.\ reports relative increase from clean performance.
    }
    \label{tab:robustness_full}
    \vspace{0.5em}
    \small
    \begin{tabular}{@{}lc|ccc|ccc|c@{}}
        \toprule
        & & \multicolumn{3}{c|}{\textbf{Depth Attenuation}} & \multicolumn{3}{c|}{\textbf{Gaussian Shadow}} & \\
        \textbf{Model} & \textbf{Original} & Low & Med & High & Low & Med & High & \textbf{Avg. Deg. $\downarrow$} \\
        \midrule
        EchoPrime & 4.87 & 5.58 & 5.71 & 5.91 & 5.55 & 5.61 & 5.78 & +16.8\% \\
        PanEcho & 5.10 & 5.10 & 5.39 & 5.46 & 5.19 & 5.21 & 5.38 & +3.7\% \\
        \midrule
        EchoMAE-L & 8.52 & 8.51 & 8.57 & 8.58 & 8.56 & 8.57 & 8.57 & +0.5\%$^\dagger$ \\
        EchoJEPA-L & 5.76 & 5.72 & 5.91 & 6.10 & 5.79 & 5.87 & 5.97 & +2.3\% \\
        EchoJEPA-G & \textbf{3.97} & \textbf{4.01} & \textbf{4.07} & \textbf{4.17} & \textbf{4.02} & \textbf{4.04} & \textbf{4.07} & \textbf{+2.3\%} \\
        \bottomrule
    \end{tabular}
    \par\vspace{0.3em}
    \small{$^\dagger$EchoMAE-L shows minimal relative degradation because its baseline (8.52 MAE) is already poor.}
\end{table*}

EchoJEPA-G maintains the best absolute performance across all perturbation levels, with MAE remaining below 4.2 even under severe degradation.
Compared to EchoPrime, EchoJEPA-G degrades by only 2.3\% on average versus 16.8\%, representing an 86\% reduction in sensitivity to acoustic artifacts.
Against PanEcho (3.7\% degradation), EchoJEPA-G shows 38\% less degradation.
This robustness gap confirms that EchoJEPA's latent prediction objective anchors features to stable anatomical structure. The controlled comparison with EchoMAE-L (identical architecture, +0.5\% degradation but from a poor 8.52 baseline) isolates the effect of pixel reconstruction, while EchoPrime's larger degradation (16.8\%) suggests that contrastive objectives also couple to acquisition-specific features.
Notably, EchoMAE-L shows minimal relative degradation (+0.5\%), but this reflects a floor effect since its baseline performance is already too poor for perturbations to meaningfully worsen.

\subsection{Multi-View Physiological Estimation}
\label{sec:exp:multiview}

Table~\ref{tab:rvsp} reports RVSP estimation, which requires integrating apical and subcostal views. EchoJEPA-G achieves 4.54 MAE on Toronto, a 17\% improvement over PanEcho.

\begin{table}[t]
    \centering
\caption{\textbf{Multi-view RVSP estimation} (MAE mmHg, lower is better). 
Predictions evaluated with multi-view probing framework.}
    \label{tab:rvsp}
    \vspace{0.5em}
    \small
    \begin{tabular}{@{}lcc@{}}
        \toprule
        \textbf{Model} & \textbf{Toronto} & \textbf{Chicago} \\
        \midrule
        EchoPrime & 5.65 & 5.29 \\
        PanEcho & 5.49 & 5.26 \\
        \midrule
        EchoMAE-L & 5.36 & 5.60 \\
        EchoJEPA-L & 5.01 & 5.05 \\
        EchoJEPA-G & \textbf{4.54} & \textbf{4.91} \\
        \bottomrule
    \end{tabular}
\end{table}

\subsection{Generalization: Adult to Pediatric Transfer}
\label{sec:exp:pediatric}

Pediatric echocardiography differs substantially from adult imaging due to smaller heart sizes, different chamber proportions, and distinct pathology distributions. We evaluate whether adult-trained representations transfer across this distribution shift on EchoNet-Pediatric (Table~\ref{tab:pediatric}).

\begin{table}[h]
    \centering
\caption{\textbf{Adult to pediatric transfer} (LVEF MAE, lower is better). 
Zero-shot: probes trained on adult data (EchoNet-Dynamic), evaluated on 
pediatric data (EchoNet-Pediatric). Fine-tuned: probes trained on 
EchoNet-Pediatric. EchoJEPA-G zero-shot outperforms all baselines even 
after probe fine-tuning on pediatric data.}
    \label{tab:pediatric}
    \vspace{0.5em}
    \small
    \begin{tabular}{@{}lcc@{}}
        \toprule
        \textbf{Model} & \textbf{Zero-Shot} & \textbf{Fine-Tuned} \\
        \midrule
        EchoPrime & 5.10 & 4.53 \\
        PanEcho & 5.66 & 5.34 \\
        \midrule
        EchoMAE-L & 6.79 & 6.75 \\
        EchoJEPA-L & 6.31 & 5.12 \\
        EchoJEPA-G & \textbf{4.32} & \textbf{3.88} \\
        \bottomrule
    \end{tabular}
\end{table}

EchoJEPA-G achieves 4.32 MAE zero-shot, 15\% lower error than EchoPrime and 36\% lower than the compute-matched reconstruction baseline. Remarkably, EchoJEPA-G without any pediatric data outperforms all baselines after fine-tuning. Fine-tuning further improves EchoJEPA-G to 3.88 MAE, establishing a new state-of-the-art for pediatric LVEF estimation. The contrast with EchoMAE-L is instructive: the reconstruction objective barely benefits from fine-tuning (6.79 $\to$ 6.75), whereas EchoJEPA-L improves substantially (6.31 $\to$ 5.12).

\begin{figure*}[t!]
    \centering
    \includegraphics[width=0.72\linewidth]{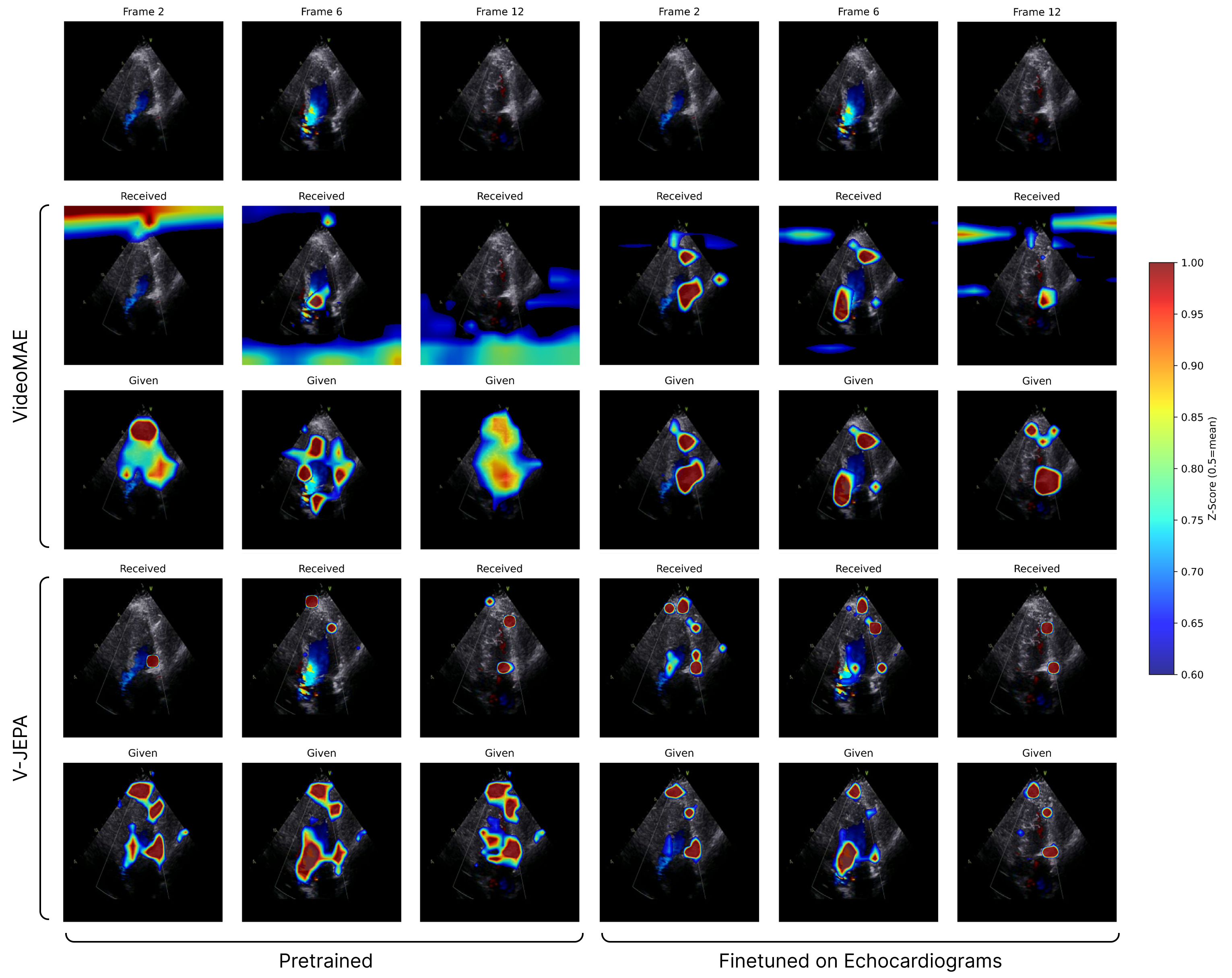}
    \caption{
        \textbf{Attention visualization comparing VideoMAE and V-JEPA.} Columns show three frames from an apical four-chamber echocardiogram under pretrained and finetuned conditions. Rows display received attention and given attention for each model. Finetuned V-JEPA in the bottom right demonstrates precise localization on valve leaflets and ventricular walls synchronized to cardiac motion.
    }
    \label{fig:attn_map}
\end{figure*}

\subsection{Clinical Interpretation}
To assess how domain-specific training reshapes learned representations, we visualize attention patterns in Figure~\ref{fig:attn_map}. VideoMAE exhibits diffuse attention across image borders and artifacts, and even after finetuning remains unfocused, tracking Doppler color intensity rather than anatomy.

EchoJEPA demonstrates anatomical localization, focusing on the mitral valve leaflets, ventricular walls, and annulus while ignoring sector background. Received attention clusters at Doppler jet edges while given attention localizes on valve structures generating flow. Across the cardiac cycle, focus shifts from valve tips during opening to chamber walls during relaxation, indicating it interprets the echocardiogram as a functional biological system.

\section{Discussion}
\label{sec:discussion}

Our results indicate that EchoJEPA achieves state-of-the-art performance across all benchmarks by shifting from pixel reconstruction to latent prediction, effectively decoupling clinical signal from acquisition artifacts.
\vspace{-5pt}

\paragraph{Objective-Domain Alignment.}
EchoJEPA outperforming compute-matched reconstruction baselines challenges the assumption that methods from natural video transfer directly to medical imaging. In natural video, texture correlates with semantics; in ultrasound, texture is largely interference noise. Pixel reconstruction forces the model to memorize this noise, while latent prediction captures only spatiotemporally coherent structures, yielding representations up to over 85\% more robust to acoustic perturbations.

\paragraph{Data Efficiency and Accessibility.}
EchoJEPA outperforming baselines with only 1\% of labeled data addresses a primary bottleneck in medical AI. Achieving strong performance with frozen backbones lowers the computational barrier, enabling clinical researchers to train lightweight probes without end-to-end fine-tuning.

\paragraph{Standardized Evaluation.}
Our multi-view probing framework resolves a crucial evaluation challenge. Prior works \citep{panecho, echoprime} relied on disparate protocols, making rigorous comparison impossible. Our framework handles variable study composition without view-specific encoders, a baseline for future research.

\paragraph{Limitations.}
Primary limitations include the reliance on proprietary data for our strongest model (though EchoJEPA-L is public), the use of synthetic rather than prospective clinical perturbations, and potential privacy risks regarding memorization that warrant further study \citep{tonekaboni2025investigationmemorizationriskhealthcare}.

\paragraph{Future Work.}
EchoJEPA opens directions including fine-grained tasks such as valve 
segmentation, prospective validation on patients with poor acoustic 
windows, integration with interpretable reasoning frameworks for 
clinical decision support, cardiac dynamics modeling 
for treatment effect prediction, and extension to other noisy modalities 
such as fetal and lung ultrasound.

\section{Conclusion}
\label{sec:conclusion}

We introduce EchoJEPA, a foundation model demonstrating that latent prediction 
outperforms pixel reconstruction for echocardiography. Trained on 18 million 
videos, EchoJEPA reduces LVEF error by 27\% over compute-matched baselines, 
achieves 78.6\% view accuracy with 1\% of labels (vs.\ 42.1\% at 100\%), 
degrades only 2.3\% under acoustic perturbations (vs.\ 16.8\%), and transfers 
zero-shot to pediatric patients better than fine-tuned models. We release 
EchoJEPA-L and our evaluation framework at \url{https://github.com/bowang-lab/EchoJEPA}.

\section*{Impact Statement}
This work aims to improve automated echocardiography analysis, which could expand access to expert-level cardiac assessment in resource-limited settings. Sample-efficient and robust models may particularly benefit patients who are currently underserved---those with obesity, lung disease, or limited access to trained cardiologists. However, automated cardiac assessment deployed without adequate validation could lead to diagnostic errors, and models trained on data from high-resource healthcare systems may underperform in other clinical contexts. We release EchoJEPA-L trained on public data to enable independent evaluation, explicitly characterize limitations, and emphasize that our models are research artifacts requiring clinical validation before deployment.

\section*{Acknowledgements}

We extend our gratitude to Amazon Web Services for providing the computational infrastructure essential to this work, and specifically to Joshua Thomas and Elizabeth Keller for their partnership in advancing AI for healthcare. We are particularly indebted to Quentin Garrido and Koustuv Sinha of the Meta AI team for their invaluable guidance on JEPA training dynamics and architectural adaptation. We also thank Augustin Toma and Jun Ma for insightful discussions regarding multi-view early fusion and cross-attention probing, as well as Zhibin Lu for his expertise in distributed systems and data pipelines. Finally, we acknowledge the University Health Network for the institutional resources that made this research possible.

\bibliography{main}
\bibliographystyle{icml2026}

\newpage
\appendix
\onecolumn

\section{Latent Space Analysis}
\label{app:latent_space}

To assess the semantic quality of learned representations, we visualize the frozen embedding spaces of all evaluated models using Uniform Manifold Approximation and Projection (UMAP). 

\begin{figure*}[h!]
    \centering
    \includegraphics[width=1.0\linewidth]{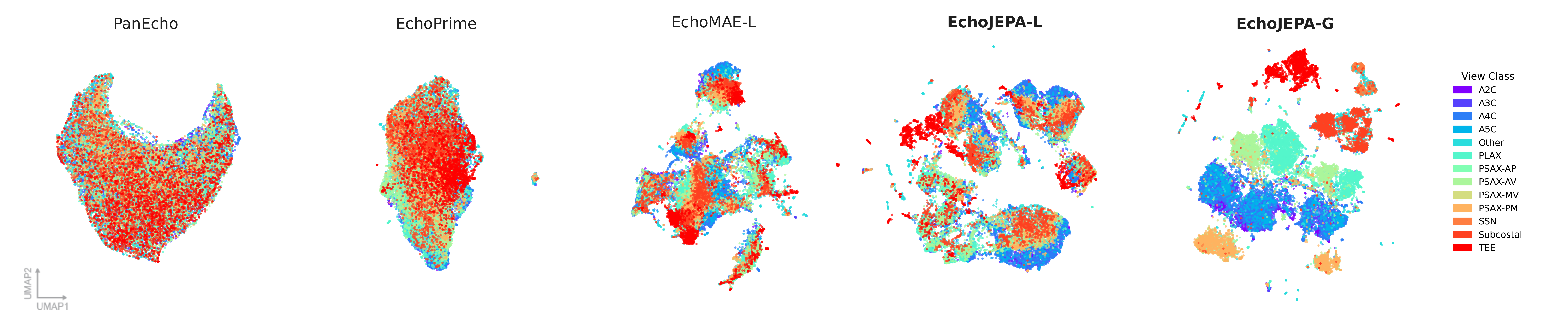}
    \caption{
        \textbf{UMAP visualization of frozen video representations colored by echocardiographic view.} Baselines (left) exhibit diffuse distributions with significant class overlap, correlating with lower probe accuracy. EchoJEPA models (right) form distinct anatomical clusters, including a clear separation of Transesophageal (TEE) views.
    }
    \label{fig:umap_views}
\end{figure*}

As shown in Figure~\ref{fig:umap_views}, \textbf{EchoJEPA-G} demonstrates strong semantic organization, forming distinct, well-separated clusters for different anatomical views (e.g., PLAX, A4C). Notably, the model clearly segregates Transesophageal (TEE) echocardiograms into a discrete cluster separate from standard Transthoracic (TTE) views. This indicates that latent prediction effectively disentangles acquisition modalities without explicit supervision.

In contrast, baselines such as \textbf{EchoMAE-L} (reconstruction), \textbf{EchoPrime} (contrastive), and \textbf{PanEcho} (supervised) exhibit diffuse embedding spaces where TTE and TEE views are largely intermixed. This suggests that alternative objectives fail to capture fundamental anatomical distinctions as effectively as latent prediction. The visual clustering quality observed here strongly correlates with the probe accuracy reported for each model.

\section{Compute-Matched Pretraining Protocol}
\label{app:compute-matched}

To isolate the effect of the pretraining objective, we train EchoJEPA-L (V-JEPA) and EchoMAE-L (VideoMAE) under a compute-matched protocol that equalizes (i) effective global batch size, (ii) total optimizer updates, (iii) input tokenization, and (iv) hardware configuration. The sole difference is the training objective: latent prediction (L1 loss on EMA encoder targets) versus pixel reconstruction (MSE loss on masked patches).

\subsection{Shared Configuration}

\paragraph{Dataset.} Both models train on MIMIC-IV-Echo~ \citep{PhysioNet-mimic-iv-echo-0.1}, comprising 525,328 echocardiogram video clips at 224$\times$224 resolution.

\paragraph{Input specification.} We sample 16 frames per clip at 8 FPS (distinct from the 24 FPS used for the proprietary EchoJEPA-G to match the standard VideoMAE protocol), tokenized with patch size 16 and tubelet size 2. Spatial augmentation uses random resized crops with scale $[0.5, 1.0]$ and aspect ratio $[0.9, 1.1]$, narrower than the V-JEPA2 defaults to preserve clinically meaningful chamber geometry (see Section~\ref{sec:method:domain}). Inputs are normalized with ImageNet statistics, tokenized with patch size 16 and tubelet size 2.

\paragraph{Compute budget.} Both runs execute on a single 8-GPU node (NVIDIA H100) with:
\begin{itemize}[nosep,leftmargin=1.5em]
    \item Effective global batch size: 1024 clips/update
    \item Total optimizer updates: 60,000 (pretraining) + 24,000 (cooldown) = 84,000
    \item Warmup updates: 12,000
\end{itemize}
This corresponds to processing approximately 86M clips and 1.4B frames per model.

\clearpage

\subsection{V-JEPA Configuration (EchoJEPA-L)}

V-JEPA parameterizes training via epochs and iterations-per-epoch (\texttt{ipe}):

\begin{table}[h]
\centering
\small
\begin{tabular}{ll}
\toprule
\textbf{Parameter} & \textbf{Value} \\
\midrule
Architecture & ViT-Large (300M params) \\
Predictor & depth=12, dim=384, heads=12 \\
Per-GPU batch & 128 $\rightarrow$ global batch = 1024 \\
Updates/epoch (\texttt{ipe}) & 300 \\
Pretraining epochs & 240 $\rightarrow$ 72,000 updates \\
Warmup epochs & 40 $\rightarrow$ 12,000 updates \\
Cooldown epochs & 80 $\rightarrow$ 24,000 updates \\
\midrule
Learning rate & $1.75 \times 10^{-4}$ (constant after warmup) \\
Final LR (cooldown) & $1.0 \times 10^{-6}$ (linear decay) \\
Weight decay & 0.04 \\
EMA momentum & 0.99925 \\
Precision & bfloat16 \\
\midrule
Masking & 8 blocks @ scale 0.15 + 2 blocks @ scale 0.7 \\
Temporal mask scale & 1.0 (full temporal extent) \\
\bottomrule
\end{tabular}
\caption{V-JEPA (EchoJEPA-L) training configuration.}
\label{tab:vjepa-config}
\end{table}

\subsection{VideoMAE Configuration (EchoMAE-L)}

VideoMAE does not natively parameterize training by total updates, so we compute epochs to match the V-JEPA budget:

\begin{table}[h]
\centering
\small
\begin{tabular}{ll}
\toprule
\textbf{Parameter} & \textbf{Value} \\
\midrule
Architecture & ViT-Large (300M params) \\
Decoder depth & 4 \\
Per-GPU batch & 32 \\
Gradient accumulation & 4 $\rightarrow$ global batch = 1024 \\
Updates/epoch & $\lfloor 525328 / 256 \rfloor / 4 = 513$ \\
Total epochs & $\lceil 84000 / 513 \rceil = 164$ \\
Warmup epochs & $\lceil 12000 / 513 \rceil = 24$ \\
\midrule
Effective Learning Rate & $3.52 \times 10^{-6}$ \\ 
Min LR & $1.0 \times 10^{-6}$ \\
Weight decay & 0.04 \\
Optimizer & AdamW, $\beta = (0.9, 0.95)$ \\
Precision & bfloat16 \\
\midrule
Masking & Tube masking, ratio = 0.9 \\
\bottomrule
\end{tabular}
\caption{VideoMAE (EchoMAE-L) training configuration.}
\label{tab:videomae-config}
\end{table}

The VideoMAE configuration executes 84,132 updates (+0.16\% relative to target), within acceptable tolerance for compute matching.

\clearpage
\subsection{Initialization}

Both models initialize from ViT-Large weights pretrained on natural video. For VideoMAE, we inflate the 2D patch embedding to 3D by replicating and normalizing across the temporal dimension, and randomly initialize the decoder. For V-JEPA, we load encoder weights directly and randomly initialize the predictor and mask tokens.


\begin{table}[h]
\centering
\small
\begin{tabular}{lcc}
\toprule
\textbf{Factor} & \textbf{EchoJEPA-L} & \textbf{EchoMAE-L} \\
\midrule
Architecture (encoder) & ViT-Large & ViT-Large \\
Parameters (encoder) & 300M & 300M \\
Training data & MIMIC-IV-Echo & MIMIC-IV-Echo \\
Training clips & 525K & 525K \\
Global batch size & 1024 & 1024 \\
Total updates & 84,000 & 84,132 \\
Input resolution & $224 \times 224 \times 16$ & $224 \times 224 \times 16$ \\
Augmentation & Matched & Matched \\
Hardware & 8$\times$H100 & 8$\times$H100 \\
\midrule
\textbf{Objective} & \textbf{Latent prediction} & \textbf{Pixel reconstruction} \\
\bottomrule
\end{tabular}
\caption{Controlled comparison summary. All factors are matched except the pretraining objective.}
\label{tab:controlled-summary}
\end{table}

\section{Ultrasound-Specific Data Augmentation}
\label{appendix:usaugment}

Training deep neural networks for ultrasound image analysis presents unique challenges due to the physics of acoustic imaging. Standard augmentation techniques (rotation, flipping, color jittering) fail to capture the domain-specific artifacts and degradations inherent to ultrasound acquisition. To address this, we employ the \texttt{usaugment} library \cite{tupper2025}, which provides physics-informed augmentation transforms specifically designed for ultrasound images.

\subsection{Overview of USAugment}

The \texttt{usaugment} library implements four ultrasound-specific augmentation transforms that simulate common artifacts and image quality variations encountered in clinical practice:

\begin{enumerate}
    \item \textbf{Depth Attenuation}: Simulates signal loss as ultrasound waves penetrate deeper tissue
    \item \textbf{Gaussian Shadow}: Simulates acoustic shadows caused by highly reflective or absorptive structures
    \item \textbf{Haze Artifact}: Simulates near-field haze and reverberation artifacts
    \item \textbf{Speckle Reduction}: Simulates varying levels of speckle filtering applied during acquisition
\end{enumerate}

These transforms integrate seamlessly with the Albumentations framework \cite{albumentations2020} and require a binary \textit{scan mask} $M \in \{0, 1\}^{H \times W}$ that identifies the active ultrasound scan region within the image frame. This mask ensures augmentations are applied only to the diagnostic region, preserving any surrounding interface elements or annotations.

\subsection{Scan Mask Generation}

For echocardiogram videos, the scan region typically appears as a sector (fan-shaped) region on a black background. We employ two strategies for scan mask generation:

\paragraph{Automatic Detection.} Given an input frame $I$, we compute a grayscale intensity image and apply a threshold $\tau$ (typically $\tau = 10$) to identify non-background pixels:
\begin{equation}
    M_{\text{raw}}(x, y) = \begin{cases}
        1 & \text{if } \bar{I}(x,y) > \tau \\
        0 & \text{otherwise}
    \end{cases}
\end{equation}
where $\bar{I}(x,y) = \frac{1}{3}\sum_{c \in \{R,G,B\}} I_c(x,y)$ is the mean intensity across color channels. The raw mask is refined using morphological closing followed by opening to remove noise and fill small holes.

\paragraph{Geometric Sector Mask.} Alternatively, we construct an idealized sector mask defined by an apex position $(x_0, y_0)$, half-angle $\theta$, and maximum radius $r_{\max}$:
\begin{equation}
    M_{\text{sector}}(x, y) = \begin{cases}
        1 & \text{if } \sqrt{(x-x_0)^2 + (y-y_0)^2} \leq r_{\max} \quad \text{and} \quad \left|\operatorname{atan2}(x-x_0, y-y_0)\right| \leq \theta \\
        0 & \text{otherwise}
    \end{cases}
\end{equation}

For batch processing of the EchoNet-Dynamic dataset, we apply augmentations to the entire frame ($M = \mathbf{1}$) since the black background regions are already at zero intensity and remain unaffected by multiplicative transforms.

\subsection{Depth Attenuation}
\label{appendix:depth_attenuation}

\subsubsection{Physical Motivation}

As ultrasound waves propagate through tissue, they undergo absorption and scattering, resulting in progressive signal attenuation with depth. This phenomenon follows the acoustic attenuation equation:
\begin{equation}
    A(z) = A_0 \cdot e^{-\alpha f z}
\end{equation}
where $A_0$ is the initial amplitude, $\alpha$ is the tissue-dependent attenuation coefficient (typically 0.5--1.0 dB/cm/MHz for soft tissue), $f$ is the ultrasound frequency, and $z$ is the propagation depth. While modern ultrasound systems apply time-gain compensation (TGC) to counteract this effect, residual depth-dependent intensity variations remain common, particularly in patients with increased body habitus or suboptimal acoustic windows.

\subsubsection{Implementation}

The depth attenuation transform generates a multiplicative attenuation map $\mathcal{A} \in [0, 1]^{H \times W}$ that decreases intensity as a function of vertical position (depth):
\begin{equation}
    \mathcal{A}(x, y) = \max\left( a_{\min}, \; 1 - \left(\frac{y}{H}\right)^{\gamma} \right)
\end{equation}
where $H$ is the image height, $\gamma$ is the attenuation rate controlling the steepness of decay, and $a_{\min}$ is the minimum attenuation factor (preventing complete signal loss). The augmented image is computed as:
\begin{equation}
    I'(x, y) = I(x, y) \cdot \mathcal{A}(x, y) \cdot M(x, y) + I(x, y) \cdot (1 - M(x, y))
\end{equation}

\subsubsection{Parameters}

\begin{table}[h]
\centering
\caption{Depth Attenuation Parameters}
\label{tab:depth_attenuation_params}
\begin{tabular}{@{}llp{7cm}@{}}
\toprule
\textbf{Parameter} & \textbf{Range} & \textbf{Description} \\
\midrule
\texttt{attenuation\_rate} ($\gamma$) & $[0.5, 3.0]$ & Controls steepness of intensity decay. Higher values produce more aggressive darkening in deep regions. \\
\texttt{max\_attenuation} ($a_{\min}$) & $[0.0, 1.0]$ & Minimum intensity multiplier at maximum depth. Setting to 0.0 allows complete signal loss; higher values preserve some visibility. \\
\texttt{p} & $[0.0, 1.0]$ & Probability of applying the transform. \\
\bottomrule
\end{tabular}
\end{table}

\subsubsection{Experimental Configurations}

For our experiments, we generate three augmented dataset variants with increasing attenuation severity:

\begin{table}[h]
\centering
\caption{Depth Attenuation Augmentation Presets}
\label{tab:depth_attenuation_presets}
\begin{tabular}{@{}lccl@{}}
\toprule
\textbf{Preset} & \textbf{$\gamma$} & \textbf{$a_{\min}$} & \textbf{Clinical Analogue} \\
\midrule
DA-075 (Mild) & 0.75 & 0.0 & Slight TGC miscalibration \\
DA-150 (Moderate) & 1.50 & 0.0 & Increased body habitus \\
DA-215 (Severe) & 2.15 & 0.0 & Poor acoustic window, obesity \\
\bottomrule
\end{tabular}
\end{table}

\subsection{Gaussian Shadow}
\label{appendix:gaussian_shadow}

\subsubsection{Physical Motivation}

Acoustic shadows occur when ultrasound waves encounter highly reflective or absorptive structures that prevent transmission to deeper tissues. In echocardiography, ribs and the sternum commonly produce acoustic shadows that partially obscure the cardiac chambers. These shadows appear as wedge-shaped or elliptical dark regions extending from the obstructing structure.

The Gaussian shadow transform, originally described by Smistad et al. \cite{smistad2018} for nerve identification in ultrasound, provides a computationally efficient approximation of these acoustic shadows using 2D Gaussian functions.

\subsubsection{Implementation}

The shadow is modeled as a localized intensity reduction centered at position $(\mu_x, \mu_y)$ with spatial extent controlled by standard deviations $(\sigma_x, \sigma_y)$:
\begin{equation}
    \mathcal{S}(x, y) = 1 - s \cdot \exp\left( -\frac{(x - \mu_x)^2}{2\sigma_x^2} - \frac{(y - \mu_y)^2}{2\sigma_y^2} \right)
\end{equation}
where $s \in [0, 1]$ is the shadow strength (maximum intensity reduction at the center). The augmented image is:
\begin{equation}
    I'(x, y) = I(x, y) \cdot \mathcal{S}(x, y)
\end{equation}

\subsubsection{Temporal Consistency for Video}

The original \texttt{usaugment} implementation randomizes the shadow position $(\mu_x, \mu_y)$ independently for each image, which is appropriate for single-frame training. However, for video-based models that leverage temporal information, this produces physically implausible flickering shadows.

We modify the implementation to generate a \textit{static} shadow map once per video, ensuring the shadow remains spatially fixed across all frames. This accurately simulates the behavior of real acoustic shadows, which maintain consistent positions relative to the transducer during a cardiac cycle (assuming the probe remains stationary).

\subsubsection{Parameters}

\begin{table}[h]
\centering
\caption{Gaussian Shadow Parameters}
\label{tab:gaussian_shadow_params}
\begin{tabular}{@{}llp{6.5cm}@{}}
\toprule
\textbf{Parameter} & \textbf{Range} & \textbf{Description} \\
\midrule
\texttt{strength} ($s$) & $[0.0, 1.0]$ & Maximum intensity reduction at shadow center. \\
\texttt{sigma\_x} ($\sigma_x/W$) & $[0.01, 0.3]$ & Horizontal extent as fraction of image width. \\
\texttt{sigma\_y} ($\sigma_y/H$) & $[0.01, 0.3]$ & Vertical extent as fraction of image height. \\
\texttt{center\_x} ($\mu_x/W$) & $[0.0, 1.0]$ & Horizontal center position (normalized). If unspecified, randomly sampled from $\mathcal{U}(0.2, 0.8)$. \\
\texttt{center\_y} ($\mu_y/H$) & $[0.0, 1.0]$ & Vertical center position (normalized). If unspecified, randomly sampled from $\mathcal{U}(0.2, 0.8)$. \\
\texttt{p} & $[0.0, 1.0]$ & Probability of applying the transform. \\
\bottomrule
\end{tabular}
\end{table}

\subsubsection{Experimental Configurations}

We define three shadow intensity presets representing varying degrees of acoustic obstruction:

\begin{table}[h]
\centering
\caption{Gaussian Shadow Augmentation Presets}
\label{tab:gaussian_shadow_presets}
\begin{tabular}{@{}lcccl@{}}
\toprule
\textbf{Preset} & \textbf{$s$} & \textbf{$\sigma_x$} & \textbf{$\sigma_y$} & \textbf{Clinical Analogue} \\
\midrule
GS-Low (Subtle) & 0.4 & 0.15 & 0.15 & Partial rib shadow, adequate window \\
GS-Med (Moderate) & 0.6 & 0.20 & 0.20 & Typical intercostal imaging \\
GS-High (Severe) & 0.8 & 0.25 & 0.25 & Significant rib/sternum obstruction \\
\bottomrule
\end{tabular}
\end{table}

\subsection{Haze Artifact}
\label{appendix:haze_artifact}

\subsubsection{Physical Motivation}

Near-field haze artifacts arise from multiple mechanisms in ultrasound imaging:
\begin{itemize}
    \item \textbf{Reverberation}: Multiple reflections between the transducer face and superficial tissue interfaces create spurious echoes that appear as diffuse brightness in the near field.
    \item \textbf{Side lobes}: Off-axis acoustic energy from the transducer elements produces low-level echoes that contaminate the main beam signal.
    \item \textbf{Clutter}: Acoustic noise from tissue motion and system electronics contributes to diffuse background signal.
\end{itemize}

These artifacts manifest as a hazy, fog-like brightness that reduces contrast in the near-field region of the image.

\subsubsection{Implementation}

The haze artifact is modeled as an additive brightness pattern concentrated near the transducer (top of image), generated using a radial gradient:
\begin{equation}
    \mathcal{H}(x, y) = h_{\max} \cdot \exp\left( -\frac{(x - x_0)^2 + (y - y_0)^2}{2\sigma_h^2} \right)
\end{equation}
where $(x_0, y_0)$ is the apex of the ultrasound sector (typically top-center), $\sigma_h$ controls the spatial extent of the haze, and $h_{\max}$ is the maximum haze intensity. The augmented image combines the original with the haze pattern:
\begin{equation}
    I'(x, y) = \min\left(1, \; I(x, y) + \mathcal{H}(x, y) \cdot M(x, y)\right)
\end{equation}

\subsubsection{Parameters}

\begin{table}[h]
\centering
\caption{Haze Artifact Parameters}
\label{tab:haze_artifact_params}
\begin{tabular}{@{}llp{7cm}@{}}
\toprule
\textbf{Parameter} & \textbf{Range} & \textbf{Description} \\
\midrule
\texttt{radius} & $[0.1, 1.0]$ & Radial extent of haze as fraction of image diagonal. \\
\texttt{sigma} ($\sigma_h$) & $[0.01, 0.2]$ & Controls the sharpness of haze falloff. Smaller values produce more concentrated haze. \\
\texttt{p} & $[0.0, 1.0]$ & Probability of applying the transform. \\
\bottomrule
\end{tabular}
\end{table}

\subsection{Speckle Reduction}
\label{appendix:speckle_reduction}

\subsubsection{Physical Motivation}

Speckle is a fundamental characteristic of coherent imaging systems, including ultrasound. It arises from constructive and destructive interference of scattered acoustic waves from sub-resolution tissue microstructure. While speckle carries tissue-specific textural information, it also reduces image contrast and obscures boundary definition.

Modern ultrasound systems offer various speckle reduction algorithms (spatial compounding, frequency compounding, adaptive filtering) with adjustable intensities. Consequently, clinical images exhibit varying levels of speckle texture depending on system settings and operator preferences.

\subsubsection{Implementation}

The speckle reduction transform applies a bilateral filter to smooth speckle while preserving edge information:
\begin{equation}
    I'(x, y) = \frac{1}{W_p} \sum_{(i,j) \in \Omega} I(i, j) \cdot \underbrace{\exp\left(-\frac{(i-x)^2 + (j-y)^2}{2\sigma_s^2}\right)}_{\text{spatial weight}} \cdot \underbrace{\exp\left(-\frac{(I(i,j) - I(x,y))^2}{2\sigma_r^2}\right)}_{\text{range weight}}
\end{equation}
where $\Omega$ is a local window centered at $(x, y)$, $\sigma_s$ is the spatial standard deviation, $\sigma_r$ is the range (intensity) standard deviation, and $W_p$ is the normalization factor.

\subsubsection{Parameters}

\begin{table}[h]
\centering
\caption{Speckle Reduction Parameters}
\label{tab:speckle_reduction_params}
\begin{tabular}{@{}llp{6.5cm}@{}}
\toprule
\textbf{Parameter} & \textbf{Range} & \textbf{Description} \\
\midrule
\texttt{sigma\_spatial} ($\sigma_s$) & $[0.1, 2.0]$ & Spatial smoothing extent. Higher values increase the effective filter radius. \\
\texttt{sigma\_color} ($\sigma_r$) & $[0.1, 2.0]$ & Range smoothing extent. Lower values preserve more edges; higher values produce more aggressive smoothing. \\
\texttt{window\_size} & $[3, 11]$ & Size of the local window (odd integer). \\
\texttt{p} & $[0.0, 1.0]$ & Probability of applying the transform. \\
\bottomrule
\end{tabular}
\end{table}

\subsection{Composing Multiple Augmentations}
\label{appendix:augmentation_composition}

The \texttt{usaugment} transforms can be composed using the Albumentations \texttt{Compose} interface to create complex, realistic degradation patterns. A typical training pipeline might apply multiple augmentations stochastically:

\begin{verbatim}
transform = A.Compose([
    DepthAttenuation(attenuation_rate=(0.5, 2.0), p=0.5),
    GaussianShadow(strength=(0.3, 0.7), sigma_x=(0.1, 0.2), p=0.3),
    HazeArtifact(radius=0.5, sigma=0.05, p=0.2),
    SpeckleReduction(sigma_spatial=0.5, sigma_color=0.5, p=0.3),
], additional_targets={"scan_mask": "mask"})
\end{verbatim}

When parameters are specified as tuples, the transform samples uniformly from the given range for each application, introducing stochastic variation across training samples.

\subsection{Augmentation Strategy for Video Models}

For video-based architectures that incorporate temporal reasoning (e.g., 3D CNNs, Video Transformers, recurrent networks), we emphasize the importance of \textit{temporal consistency} in augmentation:

\begin{enumerate}
    \item \textbf{Spatially-varying, temporally-constant augmentations} (Depth Attenuation, Gaussian Shadow): These augmentations should be computed once per video and applied identically to all frames. This preserves the temporal coherence that video models rely upon for motion estimation and maintains physical plausibility.
    
    \item \textbf{Frame-independent augmentations} (Speckle Reduction, additive noise): These may be applied independently per frame to simulate temporal variations in system noise and processing, though aggressive application may disrupt optical flow estimation.
\end{enumerate}

Our batch processing scripts implement this strategy by generating a single augmentation configuration (e.g., shadow position, attenuation map) at the start of each video and applying it consistently across all frames.

\subsection{Software and Reproducibility}
\label{appendix:software}

All ultrasound-specific augmentations are implemented using the \texttt{usaugment} library (version 1.0.0) available at \url{https://github.com/adamtupper/usaugment} and installable via PyPI:

\begin{verbatim}
pip install usaugment
\end{verbatim}

Video processing utilizes PyAV (version 12.0.0), a Pythonic binding for FFmpeg, with H.264 encoding at CRF 18 for near-lossless quality preservation. Random seeds are fixed per-video based on deterministic hashing of filenames to ensure reproducibility across runs.

Our augmentation scripts are available at \url{https://github.com/bowang-lab/EchoJEPA} and include:
\begin{itemize}
    \item \texttt{apply\_depth\_attenuation.py}: Single-video depth attenuation processing
    \item \texttt{batch\_depth\_attenuation.py}: Batch processing for depth attenuation
    \item \texttt{apply\_gaussian\_shadow.py}: Single-video Gaussian shadow processing
    \item \texttt{batch\_gaussian\_shadow.py}: Batch processing for Gaussian shadow
\end{itemize}

\section{Ablation Studies}
\label{sec:exp:ablations}

We ablate the three core components of our multi-view framework (Section~\ref{sec:method:probing}) on RVSP estimation, a task requiring integration across color Doppler A4C and PSAX-AV views.
Table~\ref{tab:ablations} reports results; we discuss each in order of impact.

\begin{table}[h]
    \centering
    \caption{\textbf{Ablation study} on RVSP estimation. Each row removes one component from the full EchoJEPA-G configuration. Relative degradation computed against baseline MAE of 4.54 mmHg.}
    \label{tab:ablations}
    \vspace{0.5em}
    \small
    \begin{tabular}{@{}lccc@{}}
        \toprule
        \textbf{Configuration} & \textbf{MAE $\downarrow$} & \textbf{$\Delta$} & \textbf{Rel.\ $\uparrow$} \\
        \midrule
        EchoJEPA-G (full) & 4.54 & -- & -- \\
        \midrule
        \quad $-$ stream embeddings & 4.63 & +0.09 & +2.0\% \\
        \quad $-$ early fusion (late avg.) & 5.09 & +0.55 & +12.1\% \\
        \quad $-$ view dropout & 5.37 & +0.83 & +18.3\% \\
        \bottomrule
    \end{tabular}
\end{table}

\paragraph{View dropout provides the largest gain (+18.3\%).}
Removing stochastic view masking during training increases MAE by 0.83 mmHg (18.3\% relative degradation).
This component, which randomly drops views with probability $p_{\text{miss}} = 0.1$, teaches the model to produce valid predictions from incomplete studies; this is precisely the scenario encountered when acoustic windows are suboptimal or acquisition time is limited.
The large effect size validates our design choice to treat missing views as a first-class concern rather than an edge case.

\paragraph{Early fusion outperforms late averaging (+12.1\%).}
Replacing early token concatenation with post-hoc prediction averaging degrades MAE by 0.55 mmHg (12.1\%).
Late fusion, as used by PanEcho, processes each view independently and combines predictions only at the output.
Early fusion enables cross-view attention from the first probe layer, allowing the model to learn which view combinations matter for RVSP, for instance, weighting the A4C tricuspid regurgitation jet against the PSAX-AV pulmonic flow.
This result confirms that multi-view reasoning requires representation-level integration, not just prediction aggregation.

\paragraph{Stream embeddings provide modest but consistent benefit (+2.0\%).}
Removing factorized stream embeddings increases MAE by 0.09 mmHg (2.0\%).
Without explicit view and clip identity, the model must infer stream membership from content alone.
The modest effect suggests that EchoJEPA's representations already encode view-discriminative features, but explicit identity injection provides a useful inductive bias.
Notably, performance remains strong without stream embeddings, indicating the framework is robust to this design choice.


\end{document}